# L'ENSEIGNEMENT DU CODE INFORMATIQUE À L'ÉCOLE
## *Prémices d'un humanisme numérique congénital*

*Mokhtar BEN HENDA*
*Chaire Unesco-ITEN, Paris*
*MICA EA-4426, Université Bordeaux Montaigne*

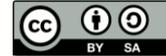

**Résumé**

Dans les systèmes éducatifs actuels, il y a un débat profond sur l'utilité de l'enseignement du code et de la programmation informatique dans les écoles. Faire évoluer l'apprentissage numérique d'une simple utilisation d'outils à la compréhension des processus du fonctionnement interne de ces outils est un débat ancien/nouveau qui a pris ses marques dans le numérique des laboratoires des années soixante, mais qui surgit de nouveau sous l'impulsion d'un numérique public à grande échelle et de nouvelles théories de l'apprentissage. Les acteurs de l'enseignement et de l'éducation discutent non seulement de la viabilité de l'enseignement du code en classe, mais aussi de certains avantages intellectuels et cognitifs pour les élèves. Le débat prend ainsi plusieurs orientations et se ressource dans l'enchevêtrement d'argumentaires et d'interprétations de tout ordre, technique, pédagogique, culturel, cognitif et psychologique. Mais sans aucun doute, ce phénomène qui augure d'une transformation profonde dans les modèles futurs de l'apprentissage et de l'enseignement, est aussi annonciateur d'un nouvel humanisme numérique quasi congénital.

**Mots-clés**
Code informatique, apprentissage, pédagogie, digital natives, humanisme numérique

**Abstract**

In today's education systems, there is a deep concern about the importance of teaching code and computer programming in schools. Moving digital learning from a simple use of tools to understanding the processes of the internal functioning of these tools is an old / new debate originated with the digital laboratories of the 1960. Today, it is emerging again under impulse of the large-scale public sphere digitalization and the new constructivist education theories. Teachers and educators discuss not only the viability of code teaching in the classroom, but also the intellectual and cognitive advantages for students. The debate thus takes several orientations and is resourced in the entanglement of arguments and interpretations of any order, technical, educational, cultural, cognitive and psychological. However, that phenomenon which undoubtedly augurs for a profound transformation in the future models of learning and teaching, is predicting a new and almost congenital digital humanism.

# Introduction

La question de l'enseignement du « code » à l'école prend de plus en plus d'importance suite à une sensibilisation grandissante aux enjeux stratégiques que ce nouveau type d'apprentissage représente pour les nouvelles générations. Le code est désormais nécessaire là où il y a une information à transformer et un système d'information à innover. Or, bien qu'un consensus grandissant semble se dessiner autour des finalités de l'enseignement du code à l'école, une ambiguïté persiste encore dans le choix d'une terminologie commune pour l'identifier. Les usages diffèrent entre « littératie numérique », « informatique », « programmation », « codage », etc., bien que tous convergent vers l'idée princeps de faire évoluer l'apprentissage numérique d'une simple utilisation des outils et des logiciels à la compréhension des processus de leur conception et de leurs modes de fonction.

La maîtrise du code devient essentielle à la compréhension d'un monde hyperconnecté qui fonctionne quasi exclusivement par des algorithmes et des processus programmatiques pour établir des identifiants de personnes et d'objets connectés, pour traiter, classer et lier entre-elles des contenues, des ressources et des supports, pour réguler et contrôler des flux et des transactions de données, pour crypter, chiffrer et compresser des données dans des systèmes sécurisés et décentralisés. Le code est de moins en moins une exclusivité des programmeurs, développeurs, experts et technologues. Il l'est encore moins des designers et des enseignants-chercheurs. En définitive, le code informatique fait exploser la bulle ésotérique des initiés et envahit progressivement la sphère publique. Il devient un objet de formation, d'enseignement et d'apprentissage aussi bien dans les écoles et les lycées que dans les collèges et les universités.

Toutes les disciplines sont désormais concernées par les théories et les techniques du code aussi bien dans les domaines scientifiques que littéraires et artistiques. Le code devient même un objet ludique de loisir et de divertissement pour les jeunes et les enfants grâce à une nouvelle génération d'outils et de dispositifs riches en interactivité et convivialité. L'apprentissage du code transforme aujourd'hui les espaces des jeux pour enfants et mobilise pour son organisation des moyens et des acteurs de la vie associative un peu partout dans le monde. Des connaissances de base en programmation, en algorithmique et en analyse computationnelle deviennent de plus en plus des critères communs aux profils métiers dans toutes les disciplines. D'où les questions qui intriguent encore bon nombre de chercheurs : les enfants et les apprenants dans les domaines des sciences humaines et sociales (SHS) devraient-ils et ont-ils intérêt à apprendre comment coder ? Si c'est le cas, ont-ils les compétences techniques et les aptitudes cognitives nécessaires pour le réussir ?

Beaucoup s'accordent sur le fait que la vraie question que tout instructeur doit d'abord se poser, tant dans le domaine des sciences humaines que dans les disciplines scientifiques et techniques, n'est pas si oui ou non les élèves doivent apprendre à coder. C'est plutôt ce qu'ils devraient apprendre en le faisant et selon quelle méthode. Vu sous cet angle, des questionnements plus précis sont à étudier : s'agit-il d'enseigner la façon algorithmique de penser aux élèves afin de mieux comprendre comment ils parviennent à agencer certaines tâches et les organiser en une série d'étapes ? Est-ce plutôt une familiarité avec les composants de base des langages de programmation, de façon à être en mesure de comprendre la sémiotique des codes et leur mode de structuration et de fonctionnement ? Est-ce, sinon, la maîtrise technique d'un langage de programmation particulier, juste pour résoudre un besoin opérationnel donné dans un domaine précis ? Ou est-ce finalement l'acquisition d'un nouveau mode opératoire des technologies numériques de nouvelle génération ? Somme toute, d'un point de vue épistémologique, en quoi le code informatique est-il différent de toute autre forme de code ou de langage au point qu'il suscite encore des réticences de son introduction dans les programmes de formation de base ou dans les sciences humaines ? C'est là une série de questions qui mériterait d'être étudiée comme anticipation à une mutation sociologique en cours qui nous conduirait à petit pas vers un nouvel humanisme numérique quasi congénital des générations à venir.

## 1 Balisons d'abord le périmètre polysémique du code

Nul doute que le terme « Code » évoque pour tous une multitude de sens qu'il serait utile de parcourir pour y dégager des liens étymologiques ou sémantiques avec le code informatique, objet de ce document.

Le Littré définit l'étymologie du terme « Code » comme un dérivé, depuis le XIIIe siècle, du terme « *codex ou caudex, assemblage de planches, des planchettes ayant servi à écrire* ». Le Codex, format primitif du livre manuscrit puis imprimé, aurait ainsi donné le sens de son support physique à différents types de contenus, depuis les recueils des lois, des constitutions, des rescrits des empereurs romains comme le « Code Théodosien » ou le « Code justinien », jusqu'aux Ordonnances ou recueils d'ordonnances des rois de France comme le « Code Michau » de 1629. Le Littré attribue aussi au Code un sens plus moderne de dispositions légales relatives à une matière spéciale ou réunies par le législateur, comme le « Code civil », et un sens figuré de « *ce qui règle dans la morale, dans les lettres, dans le goût, etc.* ».

L'étymologie du terme « Code » est donc liée à la transcription et la représentation de lois, de règles et de conventions dans un processus qui, s'il avait débuté avec le signe écrit, a progressivement migré vers d'autres formes d'abstraction sémiotique et philosophique entre l'objet et sa substance, entre l'objet et le sujet, entre le contenant et le contenu. Le code est construit au fur et à mesure que la fonction symbolique de la communication imprègne les divers aspects de la culture humaine (Dumas, et al. 2013).

Cependant, la pratique du code et de sa symbolique par l'Homme est beaucoup plus ancienne que l'acte écrit. Elle a toujours été un élément présent dans son environnement naturel, dans ses pratiques sociales et spirituelles et elle le demeurera aussi longtemps qu'il aura besoin de maîtriser la complexité de son univers d'existence. « *Durant des millénaires, c'est par la représentation que s'est manifestée la relation fondatrice sujet-objet, à laquelle chaque civilisation a donné forme et figure* […] *De même que nos ancêtres ont donné forme à leur imaginaire au cœur des grottes de la préhistoire, de même les ''primitifs du futur'' que nous sommes sont appelés, dans l'immense nébuleuse des réseaux qui se déploie, à ''dessiner'' (design) les figures-flux susceptibles de construire notre techno-imaginaire symbolique en gestation* » (Berger, Ghernaouti-Hélie, 2010:127).

Dans ce continuum historique, les enfants sont les relais naturels de cet héritage qu'il soit retransmis par un apprentissage social ou scolaire, de façon consciente ou inconsciente, volontaire ou involontaire, guidée ou spontanée. Chacun de nous possède un univers dont il a une conception propre « *même et surtout inconsciente dont il a construit une représentation qui mobilise techniques, croyances et valeurs. Toute représentation, même sommaire, s'ancre dans une culture - une société particulière - et un vécu propre* » (Naji, 2009:147). Le code est donc « *un système de sens commun aux membres d'une culture ou d'une sous-culture. Il se compose de signes et de règles ou de conventions qui déterminent comment et dans quel contexte ces signes sont utilisés et comment ils peuvent être combinés pour former des messages plus complexes* » (Fiske, 1990:19). Un code peut ainsi signifier un ensemble de lois et de règlements (code pénal), un système de signes et de symboles par lequel on traduit des informations en une série de règles de bonne conduite et de conventions dans un environnement donné (code Braille) ou une combinaison de lettres et/ou de chiffres qui autorise l'accès à certains éléments informatiques (Code ASCII).

On l'aurait donc bien compris : le concept de « convention » dans ces différentes acceptions constitue la clé de voute du bon fonctionnement de tout système de codes. C'est l'accord entre les membres d'un milieu partagé sur la façon comment les signes sont combinés entre eux pour former des codes porteurs de sens. Cette règle élémentaire s'applique autant à un code de loi qu'il faut savoir interpréter dans ses différents renvois, qu'à un code sémiotique qui donne tout son sens à un idéogramme ou un symbole, qu'à un code informatique qui ordonne les fonctionnalités exécutées par un ordinateur. Les codes peuvent ainsi être des mots ou des images, mais aussi des comportements, des signes, des gestes, des concepts et des idées. Le langage est aussi un code comme les métaphores ou les rébus. Il en va de même du système de signalisation routière, du système de classification des plantes, des algorithmes informatiques et de la nétiquette. Il n'en demeure pas moins que c'est de cette richesse polysémique du terme « Code » qu'une filiation sémantique fine traverse les différentes acceptions du concept « code » dans tous les domaines.

## 2    Le code dans la représentation du réel : de la perception de l'objet à l'abstraction du signe

La notion de « code » est une entité abstraite et dynamique qui, selon Umberto Eco, se régénère « *chaque fois qu'une séquence donnée de signes suggère, au-delà du signifié qui leur est immédiatement assignable à partir d'un système de fonction du signe, un signifié indirect* » (Eco, 1988:203). On retrouve cette notion aussi bien dans les sciences, les arts et les techniques que dans la psychologie, la philosophie, la sociologie et la communication comme un moyen d'abstraction du monde réel où « *le signifié direct s'oppose au signifié dénotatif, directement interprétable* » (Naji, 2009:147). La réalité est ainsi représentée par une abstraction progressive qui va de l'objet lui-même jusqu'à sa représentation en mots codés, équations, formules, signes, symboles, gestes ou attitudes. Michel Launey, Professeur à l'université de Paris VII, directeur du Centre d'étude des langues indigènes d'Amérique, explique ce mécanisme en soutenant le principe que « *La perception visuelle d'un objet implique un premier niveau de représentation analogique, dans les aires visuelles dites primaires (qui analysent les propriétés formelles de l'objet) et un niveau de représentation symbolique, dans les aires visuelles d'association secondaires, qui permettent la reconnaissance et la dénomination de l'objet* » (Launey, 2004:53).

Sous cet angle de perception, le code ASCII informatique serait, à titre d'exemple, une couche d'abstraction qui représente les caractères sous la forme de « 0 » et de « 1 » au-dessus de la couche inférieure qui représente les charges positives et négatives du courant électrique. Le code informatique supporterait aussi des niveaux plus élevés d'abstraction qui représentent des mots et des instructions de programmation en utilisant des lettres ASCII. Le code médical serait, lui-aussi, une abstraction des nombreux détails qui composent le dossier du patient, avec des normes établies de ce que chaque code signifie pour quelqu'un qui comprend la convention d'encodage. Précisons que le code médical standardisé est le langage utilisé dans toute l'industrie pour décrire les services médicaux en dehors du dossier écrit du patient.

Dans le domaine de la communication, le code se trouve dans le modèle de Jakobson qui désigne toutes les conventions permettant de comprendre le message de l'expéditeur par le récepteur. Comme le définit également Jacques Leclerc, spécialiste canadien en aménagement linguistique, le code en communication « *est un ensemble conventionnel de signes, soit sonores ou écrits, soit linguistiques ou non linguistiques (visuels ou autre), communs en totalité ou en partie au destinateur et au destinataire* » (Leclerc 1989:24).

Dans les arts, « *chaque production artistique accomplit ce que Roman Jakobson a nommé si intuitivement la fonction poétique de l'art visuel, soit l'enregistrement du code déjà-là, par déconstruction/construction/reconstruction continuelle d'un nouveau code. Ce travail complexe, qui est celui de l'art, fait appel et opère alors à plusieurs niveaux de signification qu'il faut circonscrire via l'analyse sémiotique* » (Carani, 1992:13). Descartes, rapporté par Jean-Pierre Cavaillé, chercheur à l'EHESS, associe cette conception à la peinture qu'il définit dans sa science de métaphysique comme l'usage d'un code de représentation ou d'une écriture qui produit de la ressemblance : « *Avec un peu d'encre, un peu de figure et de mouvement, l'artiste ou l'écrivain tracent des signes dont la lecture accomplit la fonction simulatrice et mimétique (l'imagination comme processus, construction de l'image dans la lecture, le décodage des signes)*» (Cavaillé, 1991:307).

Les sciences mathématiques constituent elle-aussi un exemple très concret de représentations sous forme d'énoncés et d'expressions contenant des théorèmes, des formules, des équations et des variables codés dans des signes et des symboles qui relèvent de l'abstraction du monde du réel. Elles nécessitent des connaissances et des compétences mathématiques pour savoir encoder et décoder, transposer, interpréter et distinguer les différentes formes de représentations d'objets réels et de situations.

Le langage naturel est lui-même « *le code culturel le plus développé et le plus finement nuancé* » confirme Gunter Kress, professeur de sémiotique à l'Université de Londres (Kress, 1988:15). On le voit bien dans le recours aux syllabaires qui constituent une économie importante de signes dans la représentation graphique de la signification. « *À chaque étape du développement de l'écriture, le nombre de symboles utiles décroît, mais comme une conséquence directe, le niveau d'abstraction des relations entre symboles écrits et signification s'accroît* » (Lecocq, 1985:354). On le voit bien aussi dans les sciences diverses qui ont affiné leur connaissance et leur représentation de leurs objets en multipliant et en épurant sans cesse les systèmes de signes par condensation, dégagement de l'essentiel, regroupement et rangement des signes dans des catégories et des ensembles plus ou moins fortement structurés pour former des codes comme le code de la route ou les langages parlés ou écrits (Chazal, 2002:80).

En définitive, il est étonnant de voir combien de codes une culture humaine peut produire. Les êtres humains organisent la vie dans des codes imbriqués : de petits codes qui s'insèrent dans des grands codes qui s'insèrent à leur tour dans d'autres codes encore plus grands et plus complexes. Les cultures alphabétiques en donnent un exemple concret en disposant les lettres selon des codes morphologiques précis pour former des mots, qui sont à leur tour ordonnés selon des codes grammaticaux en phrases, qui sont disposées par des codes sémantiques en paragraphes, qui sont arrangés par des codes de rhétorique en essais, qui sont ordonnés par des codes de performance dans une présentation ou un discours.

Ce qui est surtout étonnant, c'est que les enfants commencent depuis leur plus jeune âge à comprendre graduellement la complexité de ces codes imbriqués et les assimilent contre toute attente à des vitesses parfois déconcertantes. On a toujours vu et commenté avec surprise le processus de l'apprentissage de la langue maternelle chez l'enfant et on n'a jamais cessé de se demander sur les mécanismes langagiers qui poussent un enfant à produire des formes lexicales ou des constructions langagières qui lui sont propres sans les avoir entendu au préalable ou mimé d'un discours d'adulte.

### 3    Le *cogito* chez l'enfant vu par les théoriciens de l'apprentissage : entre la prédisposition au raisonnement et l'aptitude à l'abstraction

Le processus de l'apprentissage chez l'enfant est un champ expérimental ancien qui déborde de controverses et de théories qui se croisent, se complètent et se contredisent. Pédiatres, psychologues, sociologues, pédagogues, cogniticiens, linguistes, tous se sont penchés, chacun de son angle de perception, sur les facultés d'apprentissage de l'enfant et les processus complexes de son acquisition du langage et de ses facultés d'abstraction et de construction du raisonnement. Depuis quand l'enfant est-il un être penseur ? Dans les mêmes conditions, les enfants apprennent-ils de la même façon ? Quelles parts représentent les facultés acquises par l'expérience et l'observation comparées à celles qui sont innées ? Entre behavioristes, cognitivistes, constructivistes et socioconstructivistes, l'apprentissage chez l'enfant a été un sujet central étudié en profondeur pour donner de théories qui se croisent sur des points et se contredisent sur d'autres. Contentons-nous ici de quelques-uns parmi les plus réputés pour avoir marqué la recherche sur l'apprentissage chez les enfants.

Dans son ouvrage « Pensées sur l'éducation », John Locke (1632-1704), philosophe anglais et auteur d'une doctrine empiriste de la connaissance, développe sa théorie aristotélicienne de la « *tabula rasa* » selon laquelle l'esprit humain nait vierge de toute idées innées et que celles-ci, contrairement à l'innéisme cartésien, ne viennent que de l'expérience. L'enfant doit donc s'initier à se construire par lui-même depuis son jeune âge, avec l'aide de son entourage social (souvent ses parents ou son tuteur), en subissant les lois de la nature et du corps (Locke, 1693). Le principe lockéen de la t*abula rasa* apporte à son époque un nouvel éclairage sur l'esprit et l'intelligence de l'enfant (Morère, 2017). D'autres continueront à l'explorer comme Jean Locke, Maria Montessori, Gaston Bachelard, Jean Piaget, Lev Vigotsky, etc.

Maria Montessori (1870-1952), médecin, philosophe et psychologue, a développé une méthode d'éducation basée sur l'activité autodirigée, l'apprentissage pratique et le jeu collaboratif. Son approche se fonde sur l'idée que l'enfant est naturellement avide de connaissances et capable d'initier l'apprentissage dans un environnement d'apprentissage soutenu et bien préparé. C'est ce qu'elle appelle « l'esprit absorbant de l'enfant » (Fournier, 2017). C'est une approche qui valorise l'esprit humain et le développement de l'enfant tout entier - physique, social, émotionnel, cognitif. Les enfants font des choix créatifs dans leur apprentissage, alors que la classe et l'enseignant proposent des activités adaptées à leur âge pour guider le processus. Les enfants travaillent en groupes et individuellement pour découvrir et explorer la connaissance du monde et pour développer leur potentiel réflexif. Ces principes sont encore observés dans des méthodes pédagogiques modernes fondées sur les technologies numériques, en particulier grâce aux jeux sérieux (*serious games*), aux « classes inversées », à la dynamique des groupes et aux parcours individualisés.

Pour Gaston Bachelard (1884-1962), « *Apprendre consiste à bien identifier ses erreurs pour mieux les dépasser* », rapporte Thomas Lepeltier (2017). Cette logique est au cœur même de sa théorie sur le principe de la rupture épistémologique dans l'évolution des sciences. Toute rupture sous-tend qu'il y ait quelque chose à rompre, ce qui conduit au principe de « l'obstacle épistémologique », une résistance au développement de la connaissance au sens psychanalytique du terme que l'esprit doit surmonter en détruisant une connaissance antérieure toujours mal-faite (i.e. simple opinion) pour atteindre la connaissance scientifique. Bachelard explique aussi que ce qui est vrai de la démarche scientifique l'est aussi de l'enseignement des sciences, au sens où la notion « d'obstacle épistémologique » renvoie directement à celle « d'obstacle pédagogique ». Sur cette logique scientifique bachelardienne, Jean Piaget construira les principes pédagogiques de ses théories constructivistes de l'assimilation et de l'accommodation chez l'enfant.

Jean Piaget (1896-1980) définit sa propre recherche sur le principe d'une « épistémologie génétique » qui défend l'idée que l'enfant « *développe mentalement son apprentissage au fur et à mesure qu'il acquiert une connaissance objective du réel, un réel véritablement acquis par le raisonnement abstrait et la logique formelle, par exemple par la maitrise du raisonnement hypothético-déductif* » (Ottavi, 2017). Pour Piaget, l'enfant développe son apprentissage sur une pensée plutôt symbolique non ancrée dans l'expérience immanente mais tributaire de la réflexion et de l'abstraction aussi bien que des objets.

À ces principes constructivistes de Piaget qui aident l'enfant dans son activité d'apprentissage, Lev Vygotsky (1896-1934) apporte une conception plus ouverte sur l'environnement social. Contrairement à la conception mentaliste de Piaget selon laquelle le développement mental précède l'apprentissage individuel, Vygotsky décrit deux situations dans lesquelles l'enfant peut d'abord apprendre et accomplir seul certaines activités, mais qu'il est ensuite censé apprendre et réaliser des activités avec l'appui d'un autre pour confirmer sa capacité potentielle de développement mental. Entre ces deux situations se situe la « zone proximale de développement » (ZPD) dans laquelle l'individu peut progresser grâce à l'appui de l'autre. C'est le principe du socioconstructivisme que Vygotsky illustre dans l'une de ses citations : « Ce qu'une enfant peut réaliser aujourd'hui avec une assistance, elle peut le faire demain par elle-même » (Holowinsky, 2008:81).

Les théories de l'apprentissage sont pourtant plus diverses et parfois équivoques dès qu'elles sont confrontées à d'autres théories scientifiques, sociales ou culturelles. Ne prenant juste que la différenciation entre des situations informelles d'apprentissage social et des situations formelles d'apprentissage scientifique, dans la première, l'enfant vit en immersion dans son environnement naturel et acquiert une part importante de sa culture sans intervention délibérée de l'adulte, alors que dans la seconde, il subit un apprentissage programmé qui se ressource certes dans les connaissances sociales mais en retour les modifie et les transforme (Brossard, 2017).

Les liens dialectiques entre les différentes théories d'apprentissage parues entre le XIX et XXe siècles sont prolongés aujourd'hui dans le périmètre des technologies numériques, ce nouveau paradigme qui leur est pourtant totalement étranger. Néanmoins, cette nouvelle donne numérique est supposée avoir donné lieu à un style différent de théories d'apprentissage d'un contact plus léger que celles d'autrefois. De moins en moins de gens croient désormais qu'une seule théorie peut couvrir le champ épistémologique de l'éducation. Aucun appariement théorique particulier ne peut justifier une emprise totale d'une théorie quelconque de l'apprentissage sur l'éducation dans le monde. La théorie est désormais un assemblage hybride de propositions vérifiables et d'explications probables dérivées des manières persistantes des uns et des autres à concevoir et dispenser l'éducation et de l'effort d'abstraction dans les activités d'apprentissage des apprenants.

La question qui nous importe le plus à ce stade, est celle que nous avons posée précédemment en rapport avec ce que les enfants doivent apprendre par le numérique et surtout quelle méthode s'avère la plus pertinente pour qu'ils le fassent convenablement ? En guise de réponse à cette question, nous avons opté pour l'option encore controversée de l'apprentissage du code informatique à l'école en suivant des méthodes pédagogiques non moins controversées par beaucoup comme les jeux éducatifs, l'accompagnement social, l'apprentissage par l'erreur.

## 4   L'intéressant débat du « pour » et du « contre » relatif à l'enseignement/apprentissage du code informatique à l'école

À l'ère du numérique, dès qu'on évoque la notion du code, on converge quasi systématique vers le codage informatique alors que notre quotidien « naturel », comme nous l'avons indiqué précédemment, a toujours été un continuum d'actions de déchiffrage, d'apprentissage et d'abstraction de codes tant éthiques et déontologiques que vestimentaires et de langage, ou encore des codes bancaires et d'accès à l'Internet. Certes, dans la vie quotidienne, nous apprenons à encoder/décoder instinctivement (de façon volontaire ou involontaire sous l'emprise sécuritaire), mais dès qu'il s'agit d'une action d'enseignement et d'apprentissage programmé, nous sommes aussitôt conduits dans des situations de rupture avec le comportement « naturel » du quotidien. Pire encore quand il s'agit d'apprendre à coder en langage informatique ! Nous prenons aussitôt des postures spontanées de résistance et de rigidité parfois insurmontables (des « ruptures » et des « obstacles épistémologiques » comme les définit Gaston Bachelard). La volonté et la motivation, la prédisposition et les prérequis, mais aussi la méthode pédagogique et les conditions du contexte deviennent dès lors des paramètres essentiels pour surmonter ces obstacles et ces formes de résistance. Comment ces éléments prennent-ils forme dans le périmètre encore expérimental de l'enseignement du code informatique à l'école ? Quels sont les arguments et les contre-arguments dans ce débat qui promet d'être passionnant dans la progression de notre société vers un nouveau modèle d'humanisme numérique « congénital » ?

### 4.1   Pourquoi serait-on encore réticent à l'enseignement du code informatique à l'école ?

Enseigner le code informatique à l'école a toujours produit des résistances à différents niveaux des systèmes éducatifs. Éducateurs, pédagogues, décideurs, parents, se rejoignent souvent à négliger consciemment ou inconsciemment, l'idée de l'enseignement du code informatique aux enfants sous prétexte qu'il s'agit d'une matière (on ne savait même pas si c'était une matière scientifique ou technique) que seuls les candidats aux métiers informatiques devaient apprendre à des niveaux avancés de leur scolarité. On l'aurait donc bien compris : l'éducation est encore administrée par ceux et celles qui dans leurs propres cursus étaient formés dans une conception de spécialisation disciplinaire verticale. Le refus à toute idée d'enseignement du code informatique à l'école n'est donc pas seulement issu d'une prétendue immaturité intellectuelle des enfants mais aussi d'une classification comtienne des sciences qui aurait engendré une ségrégation arbitraire entre des esprits dits « scientifiques » et d'autres qualifiés de « littéraires » ou « artistiques ». Le code informatique, à l'instar des maths ou de la physique, serait dans la logique de cette ségrégation courante, l'apanage des initiés au raisonnement scientifique et technique et donc peu, voir nullement, accessible aux modèles de pensées dans les arts, les lettres et les SHS. Entre l'enfant et le spécialiste en SHS, chacun en fonction d'un « handicap cognitif » qui lui est attribué, se voit exclu arbitrairement de la culture du code informatique.

On comprendrait aussi la résistance à l'enseignement du code comme une contre-réaction protectionniste de la part de programmeurs. Les appels incessants et les initiatives qui émergent un peu partout pour introduire le code informatique comme matière d'enseignement depuis le primaire, constitueraient une menace pour leur avenir professionnel. Beaucoup s'en protègent d'ailleurs en évoquant la volatilité et l'obsolescence d'un tel enseignement tenant compte des langages d'encodage comme HTML, PHP, Python ou Ruby qui ont fini par périmer aussi vite que les technologies qui les utilisent. Ils ne voient donc pas pertinent d'enseigner à des enfants comment coder puisque la technologie avance rapidement. Certains sont à même de puiser dans les principes tayloriens de la division du travail pour justifier l'inutilité de l'enseignement du code à l'école comme le souligne en 2013 Chase Felker, ingénieur informatique de la société Slate.com : « *Quelle que soit la portée d'une technologie, nous n'avons pas besoin de comprendre comment cela fonctionne - notre société divise le travail afin que tous puissent utiliser les choses sans avoir l'envie de les fabriquer. Pour justifier l'apprentissage de la programmation, il faudrait que la plupart des emplois en aient besoin. Mais tout ce que je vois sont des prédictions vagues selon lesquelles la croissance des ''emplois informatiques'' signifie que nous devons soit ''programmer ou être programmé'' et que quelques entreprises riches veulent plus de programmeurs - ce qui n'est pas terriblement convaincant* » (Felker, 2013).

On peut lire aussi des positions réfractaires dans la synthèse d'une concertation nationale sur le numérique à l'école lancée par le gouvernement français le 20 janvier 2015, publiée sur le site d'Euronews le 18 février 2015 sous le titre « les enfants doivent-ils apprendre le code à l'école ? »[1]. Sous prétexte que l'encodage informatique constituerait une matière ardue et très technique pour les enfants, Régis Granarolo, président et fondateur de la MUNCI, l'association professionnelle des informaticiens de France souligne que « *ce qui peut être utile, et qui a existé il y a une vingtaine d'années, c'est l'enseignement de l'algorithmie et du pseudo-code. Mais il est inutile d'aller plus loin* ». C'est ce que défend ailleurs Johnny Castro, un expert en développement de l'enfance et membre du corps professoral du Brookhaven College à Dallas, qui dans son refus de l'enseignement du code aux enfants, appelle plutôt à les laisser jouer et profiter de l'enfance : « *Nous n'avons pas besoin de faire avancer la carrière ou l'intérêt pour les programmes informatiques jusqu'à ce que l'enfant soit plus proche de 15 ou 16 ans* » (Castro, 1999). Il croit en revanche qu'il est plus opportun pour les enfants d'explorer tous les domaines et opportunités en mathématiques, en sciences et dans d'autres domaines « avant de devenir experts dans un domaine particulier.

Jim Taylor, auteur du livre « Raising Generation Tech: Preparing Your Children for a Media-Fueled World », a écrit aussi sur ce sujet, refusant le principe de l'apprentissage anticipé du codage dans n'importe quel contexte : « *Le codage est tellement populaire parce que les parents craignent que si leurs enfants ne sont pas assez tôt sur le train technique, ils seront laissés en dehors du train pour toujours* » (Taylor, 2012). Or, contrairement à ce que les parents pensent, Taylor affirme que l'apprentissage du code à un jeune âge n'est pas la clé du succès : « *Ce qui va permettre aux enfants de réussir dans ce monde axé sur la technologie est de savoir si ils peuvent penser - de manière créative, innovante et expansive - et cela se fait par le jeu libre et non structuré* ».

---

[1] Résumé d'une concertation nationale sur le numérique à l'école lancée par le gouvernement français le 20 janvier 2015. http://fr.euronews.com/2015/02/18/enseigner-le-code-aux-enfants

En définitive, en cette deuxième décennie du XXIe siècle, le débat est à son paroxysme notamment avec l'évolution considérable de l'informatique apportée par l'intelligence artificielle, les interfaces homme-machine, la linguistique computationnelle, la réalité virtuelle et augmentée, etc. Ce qui est d'ordre à renforcer les rangs des adeptes du code informatique à l'école pour des raisons aussi bien défendables et solides que leurs contradicteurs.

### 4.2    L'enseignement du code dans la droite ligne d'un changement sociétal par le numérique

Autant les réfractaires à l'enseignement du code à l'école s'efforcent de défendre leurs positions, autant les défenseurs de cette tendance ne manquent pas non plus de ressources. Jeff Gray, professeur au département d'informatique de l'Université de l'Alabama, professeur de la Fondation Carnegie et membre du conseil consultatif de l'éducation pour le célèbre « Code.org »[2], défend l'idée qu'il y a de nombreux avantages à enseigner le code aux enfants autre que la simple justification d'apprendre à coder pour le besoin de coder. L'enseignement du codage informatique s'inscrit dans de perspectives plus ambitieuses et complexes qu'on ne pourra décrire dans les limites du présent document. On pourrait toutefois en résumer quelques-unes sous un angle pédago-technique, socioéconomique, psycho-cognitif, voire politique.

#### 4.2.1    Le code informatique : une mutation naturelle dans l'histoire des techniques

De notre point de vue, l'apprentissage du code informatique traduit d'abord un processus des plus naturels et logiques dans le courant d'une mutation historique des systèmes techniques. Bertrand Gille définit cette forme de mutation comme l'ensemble des cohérences qui se tissent à une époque donnée entre les différentes technologies et qui constituent un stade plus ou moins durable de l'évolution des techniques (Gille, 1978). L'adoption d'un système technique selon Bertrand Gille entraîne nécessairement l'adoption d'un système social correspondant afin que les cohérences soient maintenues. Notre société moderne, propulsée par un élan numérique intégral qui touche le social de l'*Homo numericus* jusqu'à son intimité profonde, ne peut maintenir une nouvelle forme de cohérence sans généraliser l'accès au code.

Nous retrouvons ce principe également dans la philosophie de Gilbert Simondon qui développe l'idée que l'invention technique ne consiste pas à fabriquer un objet à partir des principes scientifiques, mais relève plutôt d'un processus de « concrétisation » de l'objet en corrélation avec son milieu associé (Simondon, 2005). Aussi, l'apprentissage du code serait de notre point de vue la forme de concrétisation de l'objet éducatif en corrélation avec son environnement numérique. Dans le milieu technologique actuel, marqué d'un numérique pléthorique, cette corrélation ne peut être que naturelle, évidente et incontournable. La généralisation du logiciel libre et du libre accès aux codes sources, la robotique et la domotique, les techniques des puces RFID sous cutanée, la biométrie, etc. ne font que consolider ce passage inévitable vers d'autres formes d'un humanisme numérique futuriste « congénital » que nous nous hasardons de qualifier de « génétique ».

#### 4.2.2    Le code informatique, moteur d'une nouvelle société hyper-connectée

Il est d'autant plus utile de rappeler aussi que nous nous engageons dans une nouvelle forme de société dite du savoir et de la connaissance partagée, plus ancrée dans la sémantique et l'analyse automatique du sens. L'ère de la société de l'information dans laquelle techniciens, informaticiens, bibliothécaires, documentalistes, etc. ont été des assistants-relais dans l'usage du bien numérique, est en achèvement. Une nouvelle société de la connaissance et du savoir est désormais en marche exigeant que chacun joue son plein rôle de producteur, diffuseur et consommateur de ressources numériques mises en partage. Plus qu'un simple traitement de données par lequel fonctionnent les systèmes d'information conventionnels jusqu'ici, l'analyse sémantique attribue désormais un rôle central au consommateur-producteur de l'information, le seul acteur capable de déterminer sans ambigüité la densité sémantique des données qu'il produit et dont il a besoin. Pour ce faire, ce nouveau citoyen « natif digital » est appelé à disposer dès son jeune âge de compétences appropriées dont particulièrement la maîtrise du code informatique. Avec de telles compétences transversales, l'apprenant et le citoyen en général, devient un acteur producteur de ressources et administrateur de services.

---

[2] Organisation à but non lucratif et site Web éponyme comprenant des leçons gratuites de codage qui visent à encourager les gens, en particulier les étudiants des écoles aux États-Unis, à apprendre l'informatique.

Dans sa synthèse de la concertation nationale sur le numérique à l'école lancée par le gouvernement français, Marie Jamet souligne que « *Le monde est devenu numérique et nécessite des enfants formés à ce monde. Tel est l'argument le plus basique avancé et le plus partagé par les différents interlocuteurs de ce débat : il s'agit de faire des enfants des acteurs de l'environnement numérique et non plus seulement des consommateurs. L'enseignement du code doit, de ce point de vue, permettre aux enfants de comprendre les outils informatiques qui les entourent au quotidien et les entoureront dans le monde professionnel* ». C'est ce que défend aussi Laurence Bricteux, créatrice des Ateliers « Goûter du Code »[3] pour les enfants de 7 à 14 ans qui souligne que « *cela apporte aux enfants ''une vision logique'', une vue sur les jeux auxquels ils jouent et sur ce qu'il y a derrière* ».

John Naughton, universitaire britannique cité par le Guardian déclare également que « *Le monde [des enfants] va bientôt être défini par les ordinateurs : s'ils n'ont pas une meilleure compréhension de tout ça, ils seront intellectuellement paralysés. Ils grandiront, consommateurs passifs de services et d'appareils fermés, menant une vie qui sera toujours plus circonscrite par des technologies créées par une élite travaillant pour de gigantesques entreprises comme Google ou Facebook* ». Gérard Berry, informaticien professeur au Collège de France soutient aussi cette thèse : « *Il faut casser la frontière entre ceux qui sont capables de créer, et ceux qui resteront des consommateurs d'écrans* »[4].

Mitchel *Resnick* concepteur de langage de programmation à la MIT et Hadi Partovi Directeur général de « Code.org » seraient d'accord eux-aussi pour définir le codage comme une autre forme d'écriture et que chaque enfant qui apprendra à écrire ne deviendra pas forcément un romancier, ni tous ceux qui apprennent l'algèbre deviennent des mathématiciens, mais les deux sont traités comme des compétences de base que tous les enfants devraient apprendre.

Notre future société nécessiterait ainsi de nouvelles formes de médias émergents et de nouvelles formes de socialité que les générations futures seraient plus en mesure d'assimiler et de faire évoluer. Ces générations auraient cependant besoin de franchir les limites que nous lui imposons actuellement par la barrière du code pour gagner en autonomie, en liberté et surtout en innovation et créativité. Rappelons encore que la valeur réelle du codage pour les enfants n'est pas dans la maîtrise des outils, mais plutôt dans l'internalisation des méthodes d'analyse et dans les processus de pensée critique qui sont à la base d'une meilleure éducation pour créer le changement. Il ne s'agit pas seulement du code en soi autant qu'il s'agit de penser à la logique de « l'alphabétisation procédurale » qui est une capacité à réfléchir et à comprendre les processus dans le monde (Alvarenga, 2016). Compte tenu du rythme de l'innovation et de la connectivité croissante de nos appareils, de nos maisons, de nos lieux de travail et potentiellement de nos corps, il est logique de profiter de la propension naturelle des enfants à apprendre mieux et plus vite lorsqu'ils sont plus jeunes. C'est la raison pour laquelle l'enseignement du code informatique devient depuis quelques années un choix politique et stratégique.

### 4.2.3 L'enseignement du code informatique : un choix politique et stratégique

On le note bien, en effet, dans les politiques éducatives un peu partout dans le monde, aux États-Unis quand le président Barack Obama a annoncé en janvier 2016 l'initiative « Sciences informatiques pour tous » avec un budget de 4 milliards de dollars pour l'enseignement de l'informatique dans les écoles. Vingt-sept gouverneurs d'État et de nombreux chefs d'entreprise ont même signé une lettre au Congrès sur la nécessité d'un meilleur accès à l'informatique dans les programmes d'éducation K-12 (primaire et secondaire).

---

[3] Atelier de découverte de la programmation informatique pour les enfants de 7 à 14 ans, mis en place en 2015 par la franco-belge Laurence Bricteux.
[4] Signalé par Maryline Baumard dans le journal Le Monde du 23.05.2014

Au Royaume-Uni, à partir de septembre 2013, la programmation informatique à l'école est rendue obligatoire à tous les niveaux. Entre 5 et 16 ans, les enfants apprennent à coder et à utiliser des algorithmes et de langages de programmation simples pour résoudre des problèmes de calcul.

En France, les réformes scolaires ont également saisi cette opportunité pour prendre des mesures afin que l'élève soit mis en relation plus fusionnelle avec la machine, l'interface et le logiciel. Cette nouvelle donne s'est fortement traduite dans la loi n° 2013-595 du 8 juillet 2013 d'orientation et de programmation pour la refondation de l'école de la République. Il y est spécifié qu'« *Il est impératif de former les élèves à la maîtrise, avec un esprit critique, de ces outils qu'ils utilisent chaque jour dans leurs études et leurs loisirs et de permettre aux futurs citoyens de trouver leur place dans une société dont l'environnement technologique est amené à évoluer de plus en plus rapidement* ». En 2014, un rapport parlementaire[5], dirigé par les députées Laure de La Raudière et Corinne Erhel, s'est penché sur la situation du numérique en France. Parmi les points importants dans ce rapport la nécessité « *d'éveiller les élèves au code, et ce dès l'école primaire* ». La Secrétaire d'État en charge du Numérique, Axelle Lemaire, a entériné ce choix en défendant l'apprentissage du code pendant le temps périscolaire sous forme d'ateliers dédiés un peu partout en France. Elle l'a même déclaré dans une séance de l'Assemblée nationale le 14 janvier 2015 : « *Aujourd'hui, le code est partout. Il est donc essentiel que nos enfants soient autonomes dans l'environnement numérique. Pour ce faire, ils doivent apprendre ce langage qui est devenu tout aussi important qu'une langue étrangère* ».

Par ces procédures, la politique numérique française juge donc important « *que le code soit enseigné dès l'école primaire, parce que c'est un vrai outil d'autonomie pour la vie* ». Le code devient dès lors un moyen d'exercer pleinement sa citoyenneté, ce qui soutient l'idée qu'il est primordial d'aider les enfants (et les adultes chargés de leur formation) à s'approprier davantage les *modus operandi* des machines qui leur échappent.

Sur un plan stratégique, d'autres arguments de taille sont en faveur de l'enseignement du code à l'école et la préoccupation des grands acteurs politiques pour permettre à l'ensemble des enfants, d'accéder démocratiquement à cette littératie et à ses différents langages. Il s'agit d'abord d'éviter de nouvelles inégalités numériques qui risquent désormais de « *passer davantage par la capacité à coder que par le taux d'équipement des ménages* »[6]. Mais il s'agit surtout d'une question de liberté et de sécurité individuelles. Frédéric Bardeau, cofondateur de « Simplon.co », une fabrique solidaire de programmeurs, le souligne dans un billet intitulé « Le code est un vecteur d'émancipation », paru sur télérama.fr : « *Dans un texte de 2002 resté fameux et intitulé ''Code is law '' (« Le code, c'est la loi »), le juriste américain Lawrence Lessig mettait en garde contre le risque de laisser à une catégorie d'individus l'entière définition des règles et de l'architecture de l'univers numérique. Le combat mené par les hackers d'Anonymous aujourd'hui répond directement à cet impératif : savoir coder, c'est rester maître des logiciels qu'on utilise et contourner les obstacles, sur Internet, quand ceux-ci empiètent sur les libertés des utilisateurs ... Le code est un vecteur d'émancipation. L'apprendre est un acte politique : il permet de retrouver cette souveraineté numérique que l'on a abandonnée aux fabricants et aux éditeurs* »[7].

---

[5] Rapport d'information enregistré à la présidence de l'assemblée nationale le 14 mai 2014 en application de l'article 145 du règlement par la commission des affaires économiques sur le développement de l'économie numérique. https://fr.scribd.com/document/224293906/RAPPORTv6-provisoire

[6] Faut-il enseigner le code à l'école : Cap Digital et Inria ouvrent le débat. http://www.educavox.fr/accueil/debats/faut-il-enseigner-le-code-a-l-ecole-cap-digital-et-inria-ouvrent-le-debat

[7] Roch, Jean-Baptise (2014). L'algorithme dans la peau : de la nécessité d'apprendre à coder. http://www.telerama.fr/idees/code-informatique-avoir-l-algorithme-dans-la-peau,115181.php

Néanmoins, si le débat autour de l'apprentissage du code converge vers une forme de consensus qui se consolide par des initiatives d'ateliers de formation partout dans le monde[8], des textes législatifs et des réformes d'enseignement ou des recherches scientifiques, il n'en reste pas moins que l'aspect de la didactique du code reste fondamentalement importante pour que les méthodes d'apprentissage mises en place atteignent leurs objectifs. De jouer avec des animations, ce qui est souvent assimilé à une pseudo-programmation, jusqu'à la conception de jeux informatiques, l'enseignement du code dans les écoles doit passer par plusieurs étapes et plusieurs scénarios évolutifs d'apprentissage.

## 5    Une approche pédagogique graduelle pour l'apprentissage du code

Rappelons d'abord que lorsque les sciences de l'informatique ont été enseignées pour la première fois dans certaines écoles américaines et européennes dans les années 1970, c'était généralement un sujet facultatif qui se construisait sur l'enseignement de l'utilisation des logiciels sans donner un aperçu de la façon dont ils sont faits ou comment ils fonctionnent. Mais l'avènement des applications prêtes à l'emploi et des interfaces utilisateur graphiques dans les années 1980 ont marqué un virage important vers un enseignement plus évolué qui intègre des fonctions programmatiques destinées à mieux maîtriser les logiciels de production de contenus. Aujourd'hui, les macros Visual Basic, les plug-ins et les extensions, les applets Java et les scripts des feuilles de styles (CSS) sont autant de formes de programmation avec lesquels tout utilisateur débutant peut apprendre à enrichir son environnement applicatif. La technologie numérique est désormais si omniprésente qu'une appropriation plus avancée du code informatique est de plus en plus ressentie comme nécessaire et à la portée de tous. Avec des cursus de mise à niveau numérique en France comme le B2i (Brevet Informatique et Internet) et le C2i (Certificat Informatique et Internet), la culture numérique s'est bien ancrée dans le milieu scolaire et universitaire au point que l'enseignement du code informatique n'est plus considéré comme une extravagance.

Or, cet argument en faveur de l'enseignement du code est plutôt d'ordre civilisationnel lié au développement des techniques et leur généralisation dans la société. Ce qui serait plus imposant comme critère à démontrer est lié aux facultés psycho-cognitives des enfants en bas âges à assimiler une pareille évolution et l'intégrer dans leur processus cognitif d'apprentissage. Il va de soi bien entendu que ce processus ne se fera pas sans une pédagogie appropriée qui tiendrait compte à la fois de la nature même de la matière enseignée et des capacités cognitives de l'enfant pour l'intégrer.

### 5.1    L'apprentissage graduel du code selon la méthode Suzuki

À cet égard, Byrson Payne, auteur du célèbre ouvrage « Teach Your Kids to Code » et professeur d'informatique à l'Université de Géorgie du Nord aux États-Unis d'Amérique, affirme avoir appliqué sur ses enfants de 2 et 4 ans une méthode innovante développée par le japonais Shin'ichi Suzuki (1898–1998) pour enseigner du violon aux jeunes enfants en présence de leurs parents. Payne affirme à ce propos que « *l'acquisition de la musique chez l'enfant s'apparente à l'acquisition de la langue maternelle. Il doit entendre la musique, l'absorber avant de la reproduire sur son instrument, de la même façon qu'il a entendu sa langue avant de la parler* » (Byrson, 2015:XXV).

Howard Gardner, psychologue du développement américain et fondateur de la théorie des intelligences multiples, commente la méthode Suzuki et la dépeint d'une manière qui, à notre sens, s'appliquerait parfaitement à l'apprentissage du code informatique par les enfants : « *Cette méthode est très orientée vers l'apprentissage par l'oreille - probablement une décision hautement bénéfique, en considération de l'âge des enfants inscrits. On perdrait beaucoup de temps à essayer d'obtenir que*

---

[8] Lire l'article « Code Week 2016 : ateliers, sites ou applis, apprenez à coder comme vous voulez » paru dans le journal Le Monde du 14/10/2016, accessible sur http://www.lemonde.fr/campus/article/2016/10/15/code-week-2016-ateliers-sites-ou-applis-apprenez-a-coder-comme-vous-voulez_5014211_4401467.html

*les enfants en âge préscolaire lisent les notes, et l'insistance que l'on met dans de nombreux endroits à commencer sur la partition rend souvent hostiles à leurs leçons de musique des enfants qui ont par ailleurs des inclinaisons musicales* » (Gardner, 1997).

Appliquée à l'apprentissage du code informatique par les enfants, beaucoup ont considéré effectivement qu'il serait très prématuré de soumettre l'enfant au code informatique dès le début de sa scolarité. En guise de musique orale comme proposé dans le modèle Suzuki, les interfaces graphiques sous forme de jeux éducatifs pourrait servir de prélude à une immersion graduelle dans l'univers du code qui serait adaptée à l'âge de l'enfant. Dans la concertation nationale sur le numérique à l'école lancée par le gouvernement français en 2015 que nous avons citée auparavant, Colin de La Higuera, professeur d'informatique à l'université de Nantes et ex-président de la Société Informatique de France (SIF), avait proposé « *une approche ludique autour du code et de l'informatique débranchée* » au primaire, suivie d'une découverte de la programmation au collège et d'un enseignement de la science informatique et de ses grands concepts au lycée, avec toutefois des variantes selon les filières.

### 5.2   L'apprentissage ludique du code informatique : les atouts des Serious games

La conception de l'apprentissage ludique du code informatique rejoint le modèle des jeux sérieux (Serious Games) aujourd'hui au service de la pédagogie de l'enseignement à distance. Ce sont des jeux conçus pour un but au-delà du simple divertissement des jeux vidéo. Ils utilisent les leviers de motivation du design comme la concurrence, la curiosité, la collaboration et le défi individuel pour créer de la motivation et de la persévérance à dépasser les obstacles. On y retrouve tous les paramètres psycho-cognitifs du constructivisme de Piaget et du socioconstructivisme de Vygotsky que nous avons abordés auparavant. Toute connaissance est le résultat d'une expérience individuelle d'apprentissage qui fait appel aux concepts d'accommodation et d'assimilation. La construction d'un savoir, bien que personnelle, s'effectue dans ces jeux par ce que les autres apportent comme interaction.

L'un des pionniers de l'apprentissage ludique est sans doute le mathématicien et cybernéticien sud-africain Seymour Papert, collaborateur ayant accompagné Jean Piaget dans les travaux du Centre international génétique à Genève. Il fut sans conteste l'une des sources les plus originales de développement de la théorie constructiviste de l'intelligence. En 1963, il part s'installer à Boston pour travailler successivement au laboratoire d'intelligence artificielle puis au Media Lab du célèbre Massachusetts Institute of Technology (MIT), aujourd'hui leader mondial de la cybernétique. Il y participe avec Marvin Minsky à développer un programme de recherche en intelligence artificielle sur la théorie de la société de l'esprit mais s'intéressa surtout à l'usage de l'informatique dans le processus de l'apprentissage chez l'enfant. De cet intérêt naquit en 1967 son célèbre langage de programmation Logo très prisé à son temps dans le monde de l'éducation.

Le langage Logo concrétise une rencontre empirique entre le courant cognitiviste en intelligence artificielle de Minsky et le constructivisme de Jean Piaget. Il permet, dans sa principale application pédagogique, de saisir des instructions qui permettent à une tortue stylisée en forme de triangle de se déplacer en traçant des formes géométriques comme si elles sont dessinées au crayon sur du papier. Par ce principe résolument constructiviste, puisqu'il se base sur l'apprentissage par l'erreur, le jeu est clairement adapté à toutes les disciplines. À travers les instructions du langage Logo (rebaptisé Lego-Logo), les enfants construisent des processus de tout genre comme des robots avec des comportements, de la musique, des formes géométriques, etc.

Logo constitue aussi un virage important dans l'ingénierie des interfaces graphiques homme-machine. Il libère la programmation informatique pour l'enfant de la rigueur des lignes de commande, ces instructions complexes et sensibles à la casse saisies au clavier. Dans son ouvrage "In the Beginning was the Command Line", Neal Stephenson, auteur américain de science-fiction, considère l'interface graphique comme une métaphore d'abstraction entre les humains et le fonctionnement réel des dispositifs informatiques. « *Les interfaces graphiques étaient une brillante innovation de conception qui rendait les ordinateurs plus centrés sur l'homme et donc accessibles aux masses, nous conduisant vers une révolution sans précédent dans la société humaine* » (Stephenson, 1999:8).

Dans la lignée directe du Lego-Logo, l'incontournable jeu éducatif Scratch est développé en 2007 par le Groupe Lifelong Kindergarten logé dans la même enceinte du prestigieux Massachusetts Institute of Technology (MIT). Scratch est sans doute aujourd'hui l'enjeu éducatif le plus emblématique à l'échelle mondiale. C'est un langage de programmation pédagogique gratuit, conçu pour être amusant, éducatif et facile à apprendre. Il a les outils pour créer des histoires interactives, des jeux, de l'art, des simulations, et plus encore, en utilisant des programmes basés sur des blocs que l'enfant fait glisser en les attachant les unes aux autres comme un puzzle. Les structures de blocs multiples sont appelées « scripts ». Cette méthode de programmation est appelée «programmation de glisser-déposer». Elle développe chez l'enfant la pensée logique pour les aider à réussir dans d'autres activités au jour le jour et à l'école.

Des dizaines de jeux similaires ont vu le jour : Hopscoth, Tynker, Move the turtle, Alice, Pencil code, etc., tous considérés comme des moyens simples d'apprendre et d'enseigner la programmation de façon intuitive et interactive en employant un encodage visuel basé sur le glisser-déposer sans rédiger aucune ligne de code. L'un des plus connus pour avoir su employer les fonctions ludiques de Scratch à partir de la maternelle, est Kodable, développé par Grechen Huebner et Jon Mattingly dans leur startup SurfScore à Louisville dans le Kentucky. Les auteurs du jeu ont commencé par une recherche sur le développement cognitif de l'enfant afin de déterminer l'âge auquel il commence à penser logiquement et à saisir les concepts derrière la programmation. Leur idée était d'amener les enfants à se connecter tôt avec une application qui enseigne les fondements de la programmation en utilisant des symboles graphiques au lieu d'une syntaxe textuelle pour qu'ils puissent apprendre la programmation sans devoir lire ni écrire.

Le jeu présente les concepts de base de la programmation allant du simple séquençage, aux expressions de conditions (si/alors/sinon), au bouclage, aux fonctions et variables jusqu'aux commandes de la programmation orientée objet. Dans le jeu, les enfants contrôlent les déplacements des personnages appelés Fuzz, par des commandes qui les guident à travers un labyrinthe. Cela leur permet de réfléchir à un problème de plusieurs façons avant de décider d'une solution. Ils sont récompensés quand ils choisissent le meilleur chemin. Tout se fait d'une façon assez simple et abstraite pour que les jeunes enfants puissent comprendre et rester motivés au jeu.

C'est pour dire finalement que cette vague de jeux ludiques pour l'apprentissage du code informatique aux enfants de manière progressive vient remettre en cause une fausse conception selon laquelle le code et ses abstractions sont inaccessibles au mental d'un enfant. Pourtant, bien de logiques et preuves tangibles prouvent le contraire.

### 5.3 Pourquoi l'apprentissage du code n'est pas un exotisme ?

Comme nous l'avons déjà signalé, l'accès au code par les enfants est dans la droite ligne de l'ouverture du monde numérique au grand public. On n'aurait pas imaginé du temps du Web statique (Web 1.0) des années 1980 et 1990, qu'au-delà des informaticiens et des administrateurs de sites, le grand public aurait un jour la main pour produire et mettre tout seul du contenu en ligne. Pourtant, avec le Web 2.0 et le Web 3.0, chacun est désormais capable de créer du Web dynamique et collaboratif, enrichi d'applettes sophistiquées et de paramétrages complexes. Le Web, comme système technique, avait alors franchi les portes des laboratoires pour s'adapter à un nouvel environnement technologique comme un bien social accessible pour tous. La question qui s'est posée dès lors : pourquoi le code informatique ne constituerait-il pas l'étape suivante de l'accès public à la technologie numérique à l'instar d'autres langages comme HTML et XML, d'autant plus qu'avec des interfaces intuitives et conviviales et des éditeurs graphiques, chacun peut désormais produire des bouts de code en Visual basic, Java, JavaScript, Flash, CSS, etc. C'est une étape que nous envisageons voir arriver un jour, non pas pour faire de tout le monde des programmeurs nés, mais pour disposer d'outils et d'instruments intuitifs capables de générer du code à la demande. On le fait déjà avec les écrans tactiles des tablettes et des smartphones, mais nous aurons un champ de programmation autonome encore plus vaste.

Pour revenir à la notion de l'abstraction du code et du signe, il est intéressant d'opposer la rigidité supposée du code informatique à celle d'autres types de signes comme les symboles mathématiques ou physiques dont l'apprentissage est pourtant unanimement admis et encouragé. L'enfant apprend les maths et la physique dès son jeune âge et il y est encouragé par sa famille, ses enseignants et le système éducatif entier. Ce sont même des filières nobles qui constituent des choix d'orientation scolaire que tout parent rêve pour son enfant. Par contre, l'informatique reste encore rebutée sous prétexte qu'elle est encore rigide pour un enfant. Or, à notre propre entendement, un programme informatique ou un problème mathématique représenteraient les mêmes niveaux de complexité avec toutefois une relative simplicité en faveur du code informatique qui dans sa majeure partie se compose de termes et d'expressions issus du langage naturel. Les formules mathématiques sont quasi exclusivement des signes et des symboles dotés d'une charge d'abstraction plus élevée qui nécessiterait une capacité de déchiffrement plus robuste. Chaque langage nécessite néanmoins une culture d'ancrage et un univers de représentation qui mobilisent des capacités intellectuelles et cognitives appropriés chez l'enfant.

Nous inspirant d'Umberto Eco sur ce sujet, pour comprendre la logique d'une boucle en langage C++ (while/do … while/for) ou bien la théorie de l'information de Claude Shannon $H=-\sum p(x)\log p(x)$, « tout individu possède un univers dont il a une idée (même et surtout très inconsciente), dont il a construit une représentation qui mobilise techniques, croyances et valeurs. Toute représentation, même sommaire, s'ancre dans une culture - une société particulière - et un vécu propre » (Umberto Eco).

Si l'enfant réussit en général son apprentissage des mathématiques, c'est parce qu'il y a été exposé dès son jeune âge. Son cadre social en a fait une culture de référence implicite depuis des siècles. Sa complexité n'est donc pas un obstacle et l'enfant s'y adapte depuis sa naissance en commençant par la numératie, puis l'algèbre et les maths. Pour l'apprentissage du code informatique, l'idée serait semblable : procéder par un apprentissage graduel par les jeux, commençant par les notions élémentaires de l'algorithmie puis monter en complexité vers des fonctions qui nécessitent une syntaxe de programmation plus évoluée. Il n'y aurait ainsi aucun obstacle pour intégrer le code informatique dans les écoles à condition que ce soit réalisé selon une démarche qui tiendrait compte des niveaux de complexité pour chaque étape (tranche d'âge ou niveau scolaire) et des outils à base d'interfaces graphiques qui donneraient progressivement accès au code, le tout construit dans un environnement de jeu éducatif et de découverte.

## 6  Conclusion

Nous avons défendu dans ce document le besoin d'enseigner le code informatique dans les écoles pour la simple raison qu'il s'inscrit pour nous dans la droite ligne d'une prospective technologique d'un humanisme numérique en évolution permanente. Enseigner le code informatique à l'école c'est se libérer de l'hybridation analogique-numérique qui a marqué notre propre formation, et préparer une nouvelle génération de *Digital natives* plus enracinée dans une culture d'avenir que nous n'avons pas le droit de freiner ou ralentit. À des moments différents et à des endroits différents, nous et nos parents avons appris à cultiver des légumes, à construire une maison, à forger une épée ou à souffler un verre délicat, à faire du pain, à créer un soufflé, à écrire une histoire ou à tirer des cerceaux. Maintenant, c'est à notre tour d'enseigner à nos enfants quelque chose qui correspond à leur avenir : le code informatique qui sera la clé de leur monde comme le livre l'était du notre. Nous devons leur apprendre à coder, sans que ce soit une fin en soi, mais plutôt comme un moyen qui les préparerait aux transformations qui auraient modifié leur environnement de vie lorsqu'ils seront actifs et productifs. Notre monde se transforme rapidement et tant de choses que nous avons déjà produites avec la mécanique et l'industrie comme le feu et le fer, ou des outils tels que le crayon et le papier, sont maintenant conçus et produits par et pour le code.

Présenter des programmes informatiques aux enfants peut être un défi, en particulier pour ceux qui ne connaissent pas les nuances du code. Heureusement, au cours des dernières années, un certain nombre d'applications, de logiciels et de guides ont été produits, ce qui rend le sujet du codage informatique qui a été souvent complexe pour beaucoup d'entre nous, une chose évidente pour les générations futures.

Comme du temps de la Renaissance, de la révolution industrielle ou de l'ère informatique, il y aura toujours des réticences et de la résistance au changement, mais comme l'affirme si bien Alexandre Vialatte le romancier et chroniqueur français né en 1901 « Rien n'arrête le progrès. Il s'arrête tout seul ».

## 7    BIBLIOGRAPHIE